# High power spatial single-mode quantum cascade lasers at 8.9 µm

S.Forget, C.Faugeras, J-Y.Bengloan, M.Calligaro, O.Parillaud, M.Giovannini, J.Faist and C.Sirtori

*Abstract : High performance of InP-based quantum cascade lasers emitting at λ ≈ 9µm are reported. Thick electroplated gold layer was deposited on top of the laser to improve heat dissipation. With one facet high reflection coated, the devices produce a maximum output power of 175mW at 40% duty cycle at room temperature and continuous-wave operation up to 278K.*

*Introduction* : Since their first demonstration in 1994 [1], the quantum cascade laser [2] (QCL) is one of the most promising mid- and far-infrared wavelength coherent light source. Its performances in terms of power and operating temperature have been continuously improved and the range of available wavelengths extended. These unipolar lasers are able to cover the wavelength range from 3 to more than 100 µm, and more specifically operate efficiently in the two atmospheric windows (3-5 µm and 8-14µm). Those mid-IR lasers are needed for chemical sensing, free-space optical communications or spectroscopy [3, 4]. These applications require high power, room temperature continuous wave (CW) operation with a diffraction limited laser beam. We report on devices based on InP substrates with a GaInAs/AlInAs active region have proved to be the most efficient material system with regard to the emitted power and CW operation [5].

The main difficulty on the route to high power and CW operation at room temperature is the large amount of heat that must be dissipated in the device. Typical operating voltages for InP-based QCLs are 7-10 V and threshold currents may be of the order of 1 A. As the QCL's wall-plug efficiency is usually only around a few percents, almost all the injected electrical power is converted into heat in the device. Consequently, more than 10W of power must be dissipated to prevent the active region from heating, which lowers the quantum efficiency. We report in this paper spatial single-mode high average power operation of a 8.9 µm emitting InP-based QCL laser.

*Structure of the laser* : The structure of the investigated quantum cascade laser was described in full details in Ref 6. The active region of the devices consists of a four quantum well double phonon resonance design. The lattice-matched InGaAs/InAlAs quantum wells and barrier layers as well as the inner InGaAs waveguide layers were grown by molecular beam epitaxy (MBE) on a n-doped InP substrate. Metalorganic vapor-phase epitaxy (MOVPE) was used to grow the top InP cladding and contact layers. The laser design was made with a view of optimizing the output average power. The ridge waveguide size was then chosen relatively large (23 µm) to minimize optical losses. As shown in figure 1, the ridge was defined by two wet chemical etching made trenches. A thin layer of silicon nitride was deposited for insulation with an opened window on top of the ridge for current injection. A thick (20 µm) layer of electroplated gold was then deposited on the top of the laser to efficiently conduct the heat generated in the active layer [7]. Special care was taken to be sure than the Gold layer fills the trenches (see fig. 1).

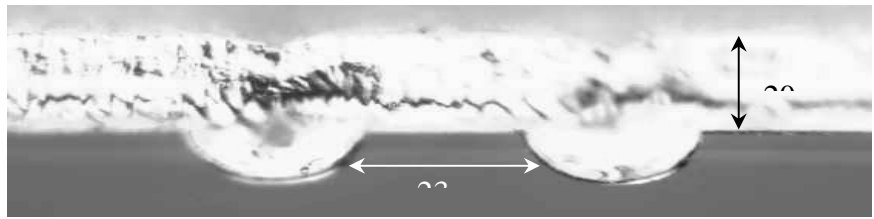

*Fig. 1 : Microscope image of the laser, with 23 µm-wide ridge and 20 µm-thick electroplated gold layer.*

After growth and processing, the lasers were cleaved into 1 and 2 mm long cavities and one of the laser facet was coated with $Al_2O_3$ and Au to realize a high reflectivity mirror (The $Al_2O_3$ layer is for electrical insulation). The lasers were then In-soldered epi-layer down on a gold coated copper mount. The emission was collected with a Ge lens (f/0.8) and focused onto the detector using a ZnSe lens (f/1). A room temperature calibrated Mercury Cadmium Telluride (MCT) detector was used for pulse measurements and a thermopile for the average power measurement.

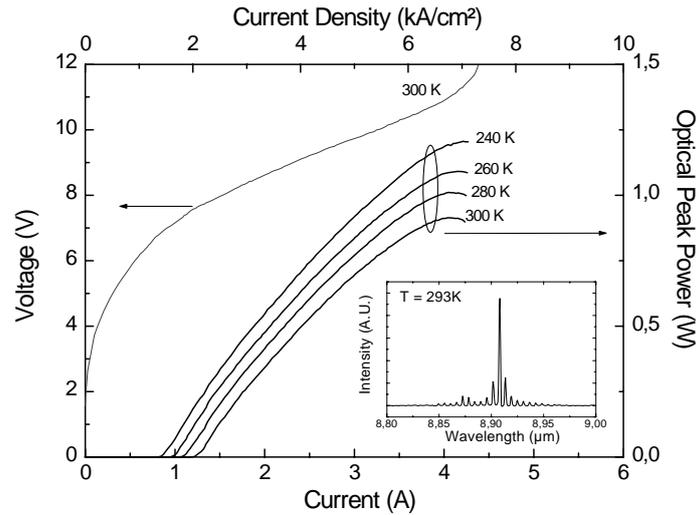

*Fig.2 : Optical power and voltage as a function of the injected current for different temperatures (2mm-long device). The exponential relation between the thresholds and the temperature led to a characteristic temperature $T_0$ of 168 K. Inset : emitting spectrum near threshold at room temperature.*

The measured spectrum of the lasers is shown in the insert of the figure 2. The central wavelength is 8.9 µm (1122.5 cm$^{-1}$) at room temperature

*Experimental results* : The lasers were first tested in pulsed mode. The typical values of the current pulses used are 100 ns with a repetition rate of 5 kHz. The temperature was varied between 77 and 300 K. All the reported power measurements represent the optical power impinging on the detector, without any correction due to the collection efficiency and the transmission coefficient of the optics (estimated around 60 %).

We present on the figure 2 the light-current (peak power) characteristics for different temperatures as well as a typical V-I curve measured at 300 K for a 2mm device. The characteristics are typical for this type of lasers with a threshold voltage around 8V. The misalignment of the bandstructure is observed at 11V where the differential resistance increases and a decrease in the outpower is observed.

We measured on the linear part of this last characteristic a resistance of only 1.1 Ω, which is evidence of good electrical properties of the laser. We obtained a maximum output peak power of 1.4 W at 77 K and still more than 900 mW at 300 K for the 2 mm lasers. The wall-plug efficiency was 3.7% at 200K and 2.4% at 300K. The 1 mm devices exhibited 580 mW of peak power at room temperature. The current threshold at 300 K was 2.5 kA/cm² for the 2 mm long lasers and 3.65 kA/cm² for the 1 mm long lasers. The corresponding slope efficiencies are 440 mW/A and 530 mW/A respectively.

The variation of the threshold current versus temperature was measured under both pulsed and CW operation. By fitting the curve obtained under pulsed operation with a usual exponential function $J_{th}=J_0 \exp(T/T_0)$ we obtained a characteristic temperature $T_0$ of 168 K. The thermal resistance at threshold could also be deduced from these measurements and was found to be around 8 K/W.

For average power characterization, the lasers were mounted on a thermoelectric cooler to monitor the temperature. The pulse duration used for this experiment was 188 ns and the repetition rate was varied from 100 kHz to 3.2 MHz. Figure 3 shows the average output power as a function of the duty cycle at a temperature of 293 K for 2 mm long lasers. The maximum output power was 175 mW for a duty cycle around 40%, which is to the best of our knowledge the highest room temperature average power reported for a QCL at this wavelength [7]. The same measurement have been done with the 1 mm laser, obtaining a maximum of about 100 mW at the optimum duty cycle of 40%.

The 2 mm long lasers operated in continuous-wave up to 278 K, with an optical power of a few mW. The power under CW operation at 77K was 417 mW and 38 mW at 260 K.

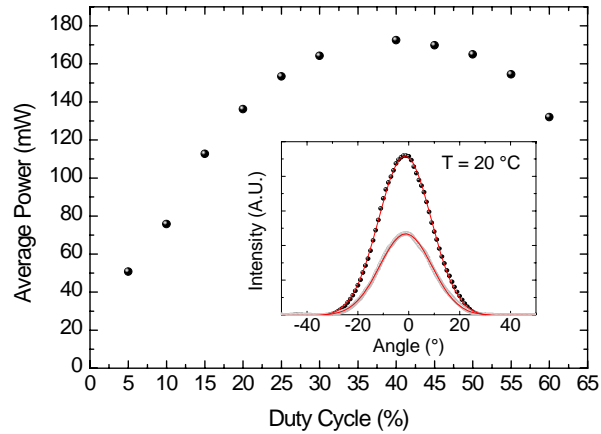

*Fig.3 : Average power at room temperature as a function of duty cycle for the 2mm-long devices. Data are measured without correcting for ~ 60% collection efficiency. Inset : far-field divergence of the laser beam in the direction parallel to the layers for two level of injected current (near threshold and at maximum power). Dots are experimental values and solid lines are the corresponding Gaussian fit.*

We also measured the far-field power distribution of the lasers in pulsed mode at room temperature. Both 1 mm and 2 mm lasers are spatially single-mode with a full angle divergence of 24° in the direction parallel to the layers, as shown in figure 3 for two different levels of injection current (just above the threshold and at maximum power). The divergence in the vertical direction is around 70°.

*Conclusion* : In conclusion, QCLs emitting at 8.9µm with 175 mW average power in spatial single-mode operation at room temperature are reported. These devices are able to perform continuous wave operation up to 278 K. These excellent optical characteristics are due to the addition of a thick Au layer which allows efficient heat evacuation and consequently enhanced performances. Further improvement is expected from narrowing of the ridge and the use of longer optical cavities.

***References :***
1. Faist, J., Capasso, F., Sivco, D.L., Sirtori, C., Hutchinson, A.L., Cho, A.Y.: "Quantum cascade laser", *Science* 264, 553 (1994).
2. Faist, J., Sirtori, C.: "InP and GaAs-based Quantum Cascade Lasers", in *Long Wavelength Infrared Semiconductor Lasers*, edited by H.K. Choi (J.Wiley and Sons, Hoboken, N.J., 2004)
3. Kosterev, A. A., Curl, R. F., Tittel, F. K., Gmachl, C., Capasso, F., Sivco, D. L., Baillargeon, J. N., Hutchinson, A. L., Cho, A. Y.: "Methane concentration and isotopic composition measurements with a mid-inf rared quantum-cascade laser", *Optics Letters* 24, pp.1762 (1999).
4. Weidmann, D., Wysocki, G., Oppenheimer, C., Tittel, F.K.: "Development of a compact quantum cascade spectrometer for field measurements of $CO_2$ isotopes", *Applied Physics B* 79, pp.907 (2004).
5. Faist, J., Sirtori, C.: *Long Wavelength Infrared Semiconductor Lasers*, Chap.5, edited by H.K. Choi (J.Wiley and Sons, Hoboken, N.J., 2004)
6. Beck, M., Hofstetter, D., Aellen, T., Faist, J., Oesterle, U., Ilegems, M., Gini, E., Melchior, H.; "Continuous-Wave Operation of a Mid-infrared semiconductor Laser at Room Temperature", *Science*, vol.295, pp.301 (2002).
7. Yu, J.S., Slivken, S., Evans, A., Doris, L., Razeghi, M.: "High power CW operation of a 6 µm Quantum Cascade laser at room temperature", *Applied Physics Letters* 83, pp. 2503 (2003).

*Authors' Affiliations :*

*S.Forget, C.Faugeras, C. Sirtori* : Pôle Materiaux et Phénomènes Quantiques, Université Paris VII, 2 place Jussieu 75251 Paris, France

*J-Y. Bengloan, M.Calligaro, O.Parillaud* : Thales Research and Technology, Domaine de Corbeville, 91404 Orsay, France

*M.Giovannini, J.Faist* : Institute of Physics, University of Neuchâtel, 2000 Neuchâtel, Switzerland.